\documentclass[lettersize,journal]{IEEEtran}
\usepackage{amsmath,amsfonts}
\usepackage{algorithmic}
\usepackage{algorithm}
\usepackage{array}
\usepackage[caption=false,font=normalsize,labelfont=sf,textfont=sf]{subfig}
\usepackage{textcomp}
\usepackage{stfloats}
\usepackage{url}
\usepackage{verbatim}
\usepackage{graphicx}
\usepackage{cite}
\usepackage{csquotes}
\usepackage[nolist]{acronym}
\usepackage[dvipsnames,usenames]{color}
\usepackage[usenames,dvipsnames]{color}

\makeatletter
\renewcommand\normalsize{%
	\@setfontsize\normalsize\@xpt\@xiipt
	\abovedisplayskip 5\p@ \@plus5\p@ \@minus5\p@
	\abovedisplayshortskip \z@ \@plus3\p@
	\belowdisplayshortskip 6\p@ \@plus3\p@ \@minus3\p@
	\belowdisplayskip \abovedisplayskip
	\let\@listi\@listI}
\makeatother

\usepackage{titlesec}
\titlespacing{\section}{0pt}{3.0ex plus .0ex minus 0.ex}{.3ex plus 0.ex}
\titlespacing{\subsection}{0pt}{2.5ex plus .0ex minus 0.ex}{.3ex plus 0.ex}
\begin{document}
\begin{acronym}
\acro{5G}{the $5$-th generation wireless system}
\acro{MIMO}{multiple-input multiple-output}
\acro{LMMSE}{linear minimum mean square error}
\acro{MSE}{mean square error}
    \acro{ANV}{average number of node-visits}
	\acro{APP}{a-posteriori probability}
	\acro{ARQ}{automated repeat request}
	\acro{ASCL}{adaptive successive cancellation list}
	\acro{ASK}{amplitude-shift keying}
	\acro{AUB}{approximated union bound}
	\acro{AWGN}{additive white Gaussian noise}
	\acro{B-DMC}{binary-input discrete memoryless channel}
	\acro{BEC}{binary erasure channel}
	\acro{BER}{bit error rate}
	\acro{biAWGN}{binary-input additive white Gaussian noise}
	%\acro{BLER}{block error rate}
	\acro{bpcu}{bits per channel use}
	\acro{BPSK}{binary phase-shift keying}
    \acro{BP}{belief propagation}
	\acro{BRGC}{binary reflected Gray code}
	\acro{BSS}{binary symmetric source}
	\acro{CC}{chase combining}
	\acro{CN}{check node}
	\acro{CRC}{cyclic redundancy check}
	\acro{CSI}{channel state information}
	\acro{DE}{density evolution}
	\acro{DMC}{discrete memoryless channel}
	\acro{DMS}{discrete memoryless source}
        \acro{dRM}{dynamic RM}
	\acro{DSCF}{dynamic successive cancellation flip}
	\acro{eMBB}{enhanced mobile broadband}
	\acro{FER}{frame error rate}
	\acro{uFER}{undetected frame error rate}
	\acro{FHT}{fast Hadamard transform}
	\acro{GA}{Gaussian approximation}
	\acro{GF}{Galois field}
	\acro{HARQ}{hybrid automated repeat request}
	\acro{i.i.d.}{independent and identically distributed}
	\acro{IF}{incremental freezing}
	\acro{IR}{incremental redundancy}
	\acro{LDPC}{low-density parity-check}
    \acro{LCT}{LDPC-coded transmission}
	\acro{LFPE}{length-flexible polar extension}
	\acro{LHS}{left hand side}
	\acro{LLR}{log-likelihood ratio}
	\acro{MAP}{maximum-a-posteriori}
	\acro{MC}{Monte Carlo}
	\acro{MLC}{multilevel coding}
	\acro{MLPC}{multilevel polar coding}
	\acro{ML}{maximum-likelihood}
	\acro{MC}{metaconverse}
	\acro{PAC}{polarization-adjusted convolutional}
	\acro{PAT}{pilot-assisted transmission}
	\acro{PCM}{polar-coded modulation}
	\acro{PDF}{probability density function}
	\acro{PE}{polar extension}
	\acro{PMF}{probability mass function}
	\acro{PM}{path metric}
	\acro{PW}{polarization weight}
	\acro{QAM}{quadrature amplitude modulation}
	\acro{QPSK}{quadrature phase-shift keying}
	\acro{QUP}{quasi-uniform puncturing}
	\acro{RCU}{random-coding union}
	\acro{RHS}{right hand side}
	\acro{RM}{Reed-Muller}
	\acro{RQUP}{reversal quasi-uniform puncturing}
	\acro{RV}{random variable}
	\acro{SC-Fano}{successive cancellation Fano}
	\acro{SCOS}{successive cancellation ordered search}
	\acro{SCF}{successive cancellation flip}
	\acro{SCL}{successive cancellation list}
	\acro{SCS}{successive cancellation stack}
	\acro{SC}{successive cancellation}
	\acro{SE}{spectral efficiency}
	\acro{SISO}{single-input single-output}
	\acro{SNR}{signal-to-noise ratio}
	\acro{SP}{set partitioning}
	\acro{UB}{union bound}
	\acro{VN}{variable node}
	%\acro{EDR}{error detection rate}
\end{acronym}

\title{Code-Aided Channel Estimation in \\ LDPC-Coded MIMO Systems}

\author{	
\IEEEauthorblockN{Binghui Shi, Yongpeng Wu, \emph{Senior Member, IEEE}, Peihong Yuan, \emph{Member, IEEE},\\ Derrick Wing Kwan Ng, \emph{Fellow, IEEE}, Xiang-Gen Xia \emph{Fellow, IEEE}, and Wenjun Zhang, \emph{Fellow, IEEE}}
\thanks{B. Shi, Y. Wu, and W. Zhang are with the Department of Electronic Engineering, Shanghai Jiao Tong University, Minhang 200240, China (e-mail: zeppoe@sjtu.edu.cn; yongpeng.wu@sjtu.edu.cn; zhangwenjun@sjtu.edu.cn) (Corresponding author: Yongpeng Wu).}
\thanks{P. Yuan is with the Research Laboratory of Electronics, Massachusetts Institute of Technology, Cambridge, MA, USA (email: phyuan@mit.edu).}
\thanks{D. W. K. Ng is with the School of Electrical Engineering and Telecommunications, University of New South Wales, Sydney, NSW, Australia (e-mail: w.k.ng@unsw.edu.au).}
\thanks{X.-G. Xia is with the Department of Electrical and Computer Engineering, University of Delaware, Newark, DE 19716, USA. (e-mail: xxia@ee.udel.edu).} \vspace{-0.5cm}
}



\maketitle

\begin{abstract}
For a multiple-input multiple-output (MIMO) system with unknown channel state information (CSI), a novel low-density parity check (LDPC)-coded transmission (LCT) scheme with joint pilot and data channel estimation is proposed. To fine-tune the CSI, a method based on the constraints introduced by the coded data from an LDPC code is designed such that the MIMO detector exploits the fine-tuned CSI. For reducing the computational burden, a coordinate ascent algorithm is employed along with several approximation methods, effectively reducing the required times of MIMO detection and computational complexity to achieve a satisfying performance. Simulation results utilizing WiMAX standard LDPC codes and quadrature phase-shift keying (QPSK) modulation demonstrate gains of up to $1.3$~dB at a frame error rate (FER) of $10^{-4}$ compared to pilot-assisted transmission (PAT) over Rayleigh block-fading channels.
\end{abstract}


\section{Introduction}
\Ac{MIMO} has been widely adopted to enhance data rates and reduce error probabilities in wireless systems \cite{mimo2,1999Capacity}. The accurate knowledge of \ac{CSI} is crucial for achieving efficient communication for a \ac{MIMO} system. In practical scenarios, pilot symbols are commonly incorporated to estimate the \ac{CSI} and the estimated \ac{CSI} is treated as the actual one at the receiver. This technique is known as \ac{PAT}~\cite{1359139} with mismatched decoding~\cite{720535}.

For short block lengths, \ac{PAT} with mismatched decoding tends to exhibit sub-optimal performance due to the trade-off between the accuracy of \ac{CSI} and code rate \cite{8486997,4299602}. Indeed, \ac{PAT} estimates \ac{CSI} only based on the pilot symbols, and neglects the inherent channel information embeded in the data symbols. Inspired by this idea, several signal processing methods have been proposed to aid channel estimation~\cite{turbo,1608563,ced,4156391,Abuthinien,9149283,8752008,6808541,9361585}. These methods are mainly based on two principles. First, received symbols are drawn from input alphabet with known distribution. Second, received signals are derived from the known codebook of the channel coding.

Based on the first principle, iterative channel estimation and data detection was proposed~\cite{4156391,Abuthinien}, which treats the detected symbols as pilots. Furthermore, iterative channel estimation and decoding was proposed~\cite{1608563,ced}, where the soft information from decoder is utilized to obtain more accurate distribution of every symbol in the input alphabet. Besides, another effective strategy, termed data-aided channel estimation~\cite{9149283,8752008,6808541}, has been designed which allows the receiver to exploit partial data symbols as additional pilot symbols for acquiring a more accurate \ac{CSI}.

Based on the second principle, polar codes with \ac{SCL} decoding was exploited to estimate \ac{CSI} for a \ac{SISO} system~\cite{9361585}. In particular, \Ac{SCL} decoding of polar codes is able to provide soft estimation of frozen bits, which can evaluate the probability that the received signals originate from the codebook. 
%For \ac{MIMO} system, the methods based on the second principle always have a lower computational complexity compared to the methods based on the first one, since less matrix operations are required in the former.

In this work, we propose a scheme realizing code-aided channel estimation for \ac{LDPC} coded systems based on the second principle. The parity checks of a \ac{LDPC} code are efficient to measure the probability of \enquote{the received sequence is a noisy observation of a codeword}, so they are applied to improve the quality of channel estimation. Furthermore, the coordinate ascent algorithm and a \ac{LLR} approximation method are proposed such that the scheme is adoptable to \ac{MIMO} systems. We denote this scheme as \ac{LCT}. We obtain a performance gain of up to $1.3$~dB at a \ac{FER} of $10^{-4}$ compared to classic \ac{PAT} schemes exploiting the same number of pilot symbols. \Ac{LCT} also performs close to the \ac{FER} of the scheme with perfect \ac{CSI}. LCT also benefits for demodulation reference signal (DMRS) patterns which only have one DMRS symbol in each subcarrier in practical 5G systems.

This paper is organized as follows. Sec. II introduces the notation and background. Sec. III describes our code-aided channel estimation algorithm. Sec. IV shows the numerical results of the proposed method for short \ac{LDPC} codes under different conditions. Sec. V concludes the paper.

%\IEEEpubidadjcol

\section{Preliminaries}

Uppercase letters, e.g., $X$, denote random variables while lowercase letters, e.g., $x$, denote their realizations. The probability distribution of $X$ evaluated at $x$ is written as $P_X(x)$. Lower case boldface letters, e.g., $\boldsymbol{x}$, denote column vectors. Capital boldface letters, e.g., $\boldsymbol{X}$, denote  matrices and capital boldface letters with a bar over them, e.g., $\bar{\boldsymbol{X}}$, denote random matrices. Let $\boldsymbol{0}_{m \times n}$ be the all zero matrix and $\boldsymbol{E}_{i,j}$ be the matrix whose element on the $i$th row and $j$th column is 1 and all the other elements are 0. We represent $\Vert\cdot\Vert$ for the $l_2$-norm and $\Vert\cdot\Vert_{F}$ for the Frobenius norm. Let $\oplus$ denote the exclusive OR operation, $\mathbb{C}$ denote the complex field, $\Re$ denote the real part of a complex number and $(\cdot)^H$ denote the conjugate transpose.

\subsection{System Model}
Consider a point-to-point \ac{MIMO} system with $N_t$ transmit antennas and $N_r$ receive antennas. We assume that the fading coefficients in $\boldsymbol{H}$ are constant during $n$ channel uses. Let $\mathcal{X}$ denote the input alphabet and the channel output is
\begin{equation*}
	\boldsymbol{Y}= \boldsymbol{H} \boldsymbol{X}+\boldsymbol{Z},
	\label{chan}
\end{equation*}
where $\boldsymbol{X} \in \mathcal{X}^{N_t\times n}$ and $\boldsymbol{Y} \in \mathbb{C}^{N_r\times n}$ are the transmitted and received signal matrices, $\boldsymbol{H} \in \mathbb{C}^{N_r\times N_t}$ is the fading channel matrix, and $\boldsymbol{Z}\in\mathbb{C}^{N_r\times n}$ is \ac{AWGN} matrix whose entries follow \ac{i.i.d.} complex Gaussian distribution with zero-mean and variance $2\sigma^2$, i.e., $\mathcal{C N}\left(0,2 \sigma^2\right)$. Neither the transmitter nor the receiver knows the value of $\boldsymbol{H}$ or its probability distribution $\bar{\boldsymbol{H}}$. We assume that the noise variance $2\sigma^2$ is known to the receiver by long term measurement.

Consider the use of \ac{QPSK} and $16$-\ac{QAM} with Gray labeling. The input alphabet is $\mathcal{X}=\frac{1}{\sqrt{2}}\{\pm 1 \pm j \}$ for \ac{QPSK} and $\mathcal{X}=\frac{1}{\sqrt{10}}\{\pm a \pm j b|a,b\in\{1,3\}\}$ for $16$-QAM. We map a binary vector $\boldsymbol{c}\in\{0,1\}^{n N_t k}$ to $\boldsymbol{X} \in \mathcal{X}^{N_t\times n}$ through the modulator with $k=2$ for \ac{QPSK} and $k=4$ for $16$-QAM. A random interleaver permutes the encoded bits $\boldsymbol{c}$ right before the modulator. We assume that all the symbols in $\mathcal{X}$ are equally probable.

\subsection{LDPC Codes}
An \ac{LDPC} code of rate $r=K/N$ is defined by its parity check matrix $H$ of size $(N-K)\times N$ whose elements are binary, which encodes the information bits $\{u_i\}_{i=1}^K$ to codeword  $\{c_i\}_{i=1}^N$. Each row of the parity check matrix $H$ represents a constraint on the encoded message $\boldsymbol{c}$. The constraint of the $i$th row is 
$$
\bigoplus_{j=1}^{n_i} c_{k_{i,j}}=0,\quad i=1, \ldots, N-K,
$$
where $n_i$ represents the number of ones in the $i$th row of the parity check matrix $H$ and $k_{i,j}$ represents the position of the $j$th one in the $i$th row. Let $W=\underset{i=1,\ldots,M}{\max}n_i$ represent the maximum number of ones in all the rows. For ease of presentation, we denote $N-K$ as $M$ in the following.

The parity check matrix $H$ has a significant impact on the performance of an \ac{LDPC} code, so the selection of $H$ is specified in numerous standards. In this work, \ac{LDPC} codes from WiMAX in IEEE 802.16 are adopted.

The \ac{BP} decoder iteratively computes the probability $p_i$ that the bit $c_i$ in $\boldsymbol{c}$ equals 1 when all the constraints are satisfied. The iteration count $R$ of the BP decoder refers to the maximum iteration number in case the parity check matrix $H$ cannot be satisfied. %The BP decoder will stop if all the constraints are satisfied or it has run $R$ times. In practice, the number $R$ affects the performance and complexity of the decoding. More iterations may result in a better bit error rate performance but lead to a linear increase in complexity.

\subsection{Pilot-Assisted Transmission}
In \ac{PAT}, the first $n_p$ symbols are pilot symbols $\boldsymbol{X}_p$ and the remaining $n_d=n-n_p$ symbols $\boldsymbol{X}_d$ are coded. The pilot and coded symbols have the same amount of energy, and the pilot symbols are known to the receiver. The received signals of $\boldsymbol{X}_p$ and $\boldsymbol{X}_d$ are denoted as $\boldsymbol{Y}_p$ and $\boldsymbol{Y}_d$, respectively. We set $N = n_d N_t k$ to keep the block length fixed. In this paper with \ac{PAT}, a \ac{LMMSE} channel estimator is adopted to obtain CSI with the received signal $\boldsymbol{Y}_p$. The \ac{LMMSE} estimator is \cite{1193803}
$$\hat{\boldsymbol{H}}=\boldsymbol{Y}_p\boldsymbol{X}_p^H(\boldsymbol{X}_p\boldsymbol{X}_p^H+2\sigma^2\boldsymbol{I}_{N_t})^{-1}.$$

We adopt a bit-interleaved coded modulation (BICM)~\cite{bicm2} encoder at the transmitter and a bit-metric decoder at the receiver. With given $\boldsymbol{Y}_d,\hat{\boldsymbol{H}}$, a \ac{ML} detector with log-sum approximation is applied to obtain the bit-wise \acp{LLR}. Let $L_i,i=1,\ldots,N$, denote the \acp{LLR} of every bit, then for $i=1,\ldots,N$,
\begin{equation}
	L_i = \frac{1}{2\sigma^2}\left(\underset{\boldsymbol{x}\in Q_i^1}{\min}\Vert\boldsymbol{y}_i-\hat{\boldsymbol{H}}\boldsymbol{x}\Vert^2-\underset{\boldsymbol{x}\in Q_i^0}{\min}\Vert\boldsymbol{y}_i-\hat{\boldsymbol{H}}\boldsymbol{x}\Vert^2\right),
	\label{ml}
\end{equation}
where $\boldsymbol{y}_i\in \mathbb{C}^{N_r}$ is the column of $\boldsymbol{Y}_d$ which contains $c_i$ and $Q_i^0\subset \mathcal{X}^{N_t}$ contains all the vectors whose $c_i=0$, $Q_i^1\subset \mathcal{X}^{N_t}$ contains all the vectors whose $c_i=1$.

%\begin{figure}[!t]
%	\centering
%	\includegraphics[width=0.4\textwidth]{pat.pdf}
%	\caption{A PAT frame structure with $N_t$ transmit antennas. Shaded and empty boxes represent pilot and data symbols, respectively.}
%	\label{fig_n}
%\end{figure}

\section{Code-Aided Channel Estimation}
This section presents a code-aided channel estimation scheme for an LDPC-coded MIMO system. Through this scheme, we can fine-tune the channel estimation and achieve a better \ac{FER} performance.

Let binary random variables $\{Z_i\}_{i=1}^M$ denote whether the constraint of the $i$th row of the parity check matrix $H$ is satisfied, and $z_i=0$ indicates that the constraint of the $i$th row is satisfied. Let binary random variables $\{\Pi_i\}_{i=1}^M$ denote whether the constraints of the first $i$ rows of the parity check matrix $H$ are satisfied and $\Pi_i=0$ indicates that they are satisfied. For notation consistency, let $\Pi_0$ denote a binary random variable that $P_{\Pi_0}(0)=1$ and is uncorrelated to $Z_i$.

Since the parity check matrix $H$ of an LDPC code is sparse, the probability that the received sequence is a codeword can be accurately approximated by the probability that all the constraints of LDPC code are satisfied. Then, we calculate this probability with the given $\hat{\boldsymbol{H}}$ to fine-tune the channel estimation as
\begin{equation}
	\begin{aligned}
		\boldsymbol{H}^\star&=\underset{\hat{\boldsymbol{H}}}{\arg \max } P_{\Pi_M|\bar{\boldsymbol{Y}}_d,\bar{\boldsymbol{H}}}(0|\boldsymbol{Y}_d,\hat{\boldsymbol{H}}) \\
		&=\underset{\hat{\boldsymbol{H}}}{\arg \max }\prod_{i=1}^{M}P_{Z_{M-i+1}|\Pi_{M-i},\bar{\boldsymbol{Y}}_d,\bar{\boldsymbol{H}}}(0|0,\boldsymbol{Y}_d,\hat{\boldsymbol{H}}).
	\end{aligned}
	\label{prod}
\end{equation}

\subsection{Channel Estimation Fine-tuning}
With the fixed $\boldsymbol{Y}_d,\hat{\boldsymbol{H}}$, we first calculate the LLR of every bit by the ML detector \eqref{ml}. For simplicity, we ignore $\boldsymbol{Y}_d,\hat{\boldsymbol{H}}$ in the following equations.

Let $\{C_i\}_{i=1}^N$ be the binary random variables representing every received bit and $\{c_i\}_{i=1}^N$ are their realizations. Then, for $i = 1,2,\ldots,M$, we have

\begin{equation}
	\begin{aligned}
		P_{Z_i}(0)&=P\left(\bigoplus_{j=1}^{n_i} c_{k_{i,j}}=0\right)\\
		&=\frac{1}{2}+\frac{1}{2} \prod_{j=1}^{n_i}\left(1-2  P_{C_{k_{i,j}}}(1)\right).
	\end{aligned}
	\label{equal1}
\end{equation}

If all the rows of parity check matrix are uncorrelated, then every term in \eqref{prod} is
\begin{equation}
	P_{Z_{M-i+1}\mid \Pi_{M-i}}(0\mid 0)=P_{Z_{M-i+1}}(0)
	\label{noc}
\end{equation}
and R.H.S of \eqref{noc} is obtained by \eqref{equal1}.

However, since the rows with a 1 in the same column are correlated, \eqref{noc} does not always hold. According to the algorithm of BP decoder, every term of the product in \eqref{prod} is approximated as follows.

Let $L_{X\mid Y}=\ln\left(\frac{P_{X\mid Y}\left(x=0\mid y=0\right)}{P_{X\mid Y}\left(x=1\mid y=0\right)}\right)$ represent the LLR of $x$ given $y=0$ and $L_X=\ln\left(\frac{P\left(x=0\right)}{P\left(x=1\right)}\right)$ represent the \ac{LLR} of $x$. Then, convert \eqref{equal1} into \ac{LLR} form that yields
\begin{equation}
	L_{Z_i\mid \Pi_{i-1}}=2\tanh^{-1} \left(\prod_{j=1}^{n_i}\left(\tanh\left(\frac{L_{C_{k_{i,j}} \left|\Pi_{i-1}\right.}}{2}\right)\right)\right),
	\label{everyllr}
\end{equation}
where $p_{C_{j}\mid \Pi_{i}}\left(1 \mid 0\right)$ in $L_{C_{j}\left|\Pi_{i}\right.}$ is updated by the following steps and $L_{C_{j}\left|\Pi_{0}\right.}=L_j$ are the \acp{LLR} computed by the $\boldsymbol{Y}_d,\hat{\boldsymbol{H}}$. For $j=1,2,\ldots,N$, we have
\begin{equation}
	P_{C_{j}\mid \Pi_{i}}\left(0 \mid 0\right)=\frac{P_{Z_i \mid C_{j}, \Pi_{i-1}}(0\mid 0,0)}{P_{Z_i\mid \Pi_{i-1}}\left(0 \mid 0\right)} P_{C_{j}\mid \Pi_{i-1}}\left(0 \mid 0\right),
	\label{updatep}
\end{equation}
where if $j \in \{k_{i,m}\mid m=1,\ldots,n_i\},$
$$
\begin{aligned}
	P_{Z_i \mid C_{j}, \Pi_{i-1}}&\left(0\mid 0,0\right)=\frac{1}{2}\\
	&+\frac{1}{2}\prod_{m=1, k_{i,m}\neq j}^{n_i} \left(1-2 P_{C_{k_{i,m}}\mid \Pi_{i-1}}\left(1 \mid 0\right)\right),
\end{aligned}
$$
else 
$$P_{Z_i \mid C_{j}, \Pi_{i-1}}\left(0\mid 0,0\right)=P_{Z_i\mid \Pi_{i-1}}\left(0 \mid 0\right).$$

We convert \eqref{updatep} to LLR form and obtain if $j \in \{k_{i,m}\mid m=1,\ldots,n_i\},$
\begin{equation}
	\begin{aligned}
		L_{C_{j}\left|\Pi_{i}\right.}=&2\tanh^{-1}\left(\prod_{m =1,k_{i,m}\neq j}^{n_i}\left(\tanh\left(\frac{L_{C_{k_{i,m}} \left|\Pi_{i-1}\right.}}{2}\right)\right)\right)\\
		&+L_{C_{j}\left|\Pi_{i-1}\right.}.
	\end{aligned}
	\label{update}
\end{equation}

Also, we convert $p_{\Pi_{i}}=p_{Z_{i}\mid \Pi_{i-1}} p_{\Pi_{i-1}}$ to \ac{LLR} form and have
\begin{equation}
	L_{\Pi_{i}}=L_{Z_{i}\mid \Pi_{i-1}} + L_{\Pi_{i-1}} - \log\left(1+e^{L_{Z_{i}\mid \Pi_{i-1}}} + e^{L_{\Pi_{i-1}}}\right).
	\label{combine}
\end{equation}

The algorithm to calculate $P_{\Pi_M}(0)$ in \ac{LLR} form is summarized in Algorithm~\ref{cal}.

\begin{algorithm}
	%\textsl{}\setstretch{1.8}
	\renewcommand{\algorithmicrequire}{\textbf{Input:}}
	\renewcommand{\algorithmicensure}{\textbf{Output:}}
	\caption{Metric Calculation Algorithm}
	\label{cal}
	\begin{algorithmic}[1]
		\REQUIRE the LLRs of every bit $\{L_i\}_{i=1}^N$, parity check matrix $H$
		\FOR{$i=1$ to $M$}
		\STATE calculate $L_{Z_i\mid \Pi_{i-1}}$ via \eqref{everyllr}
		\STATE update $L_{C_{j}\left|\Pi_{i}\right.}$ via \eqref{update}
		\STATE combine $L_{Z_{i}\mid \Pi_{i-1}}$ and $L_{\Pi_{i-1}}$ via \eqref{combine} and get $L_{\Pi_{i}}$
		\ENDFOR
		\ENSURE $L_{\Pi_M}$, which is $P_{\Pi_M}(0)$ in LLR form
	\end{algorithmic}  
\end{algorithm}

The computational complexity of calculating $L_{\Pi_M}$ given $\boldsymbol{Y}_d, \hat{\boldsymbol{H}}$ is $\mathcal{O}(MW)$, which is the same as one round of the BP decoder. We denote Algorithm~\ref{cal} without updating $L_{C_{j}\left|\Pi_{i}\right.}$ via \eqref{update} in line 3 as Algorithm~\ref{cal}a. The comparison between Algorithm~\ref{cal} and Algorithm~\ref{cal}a is given in Section IV.

\subsection{Coordinate Ascent Algorithm}
A direct method to solve \eqref{prod} is via an exhausting search. However, the search space in \eqref{prod} grows exponentially in terms of $N_rN_t$. As a compromise approach, we adopt a coordinate ascent method to acquire a high-quality solution. A channel estimation $\hat{\boldsymbol{H}}$ is firstly obtained by \ac{LMMSE}. In the coordinate ascent, every element of $\hat{\boldsymbol{H}}$ is optimized one-by-one. When optimizing one element, its real and imaginary parts are optimized sequentially. When optimizing one part, we adopt a grid search to find the maximum point in a fixed neighborhood with the other elements fixed. Algorithm~\ref{alg1} specifies the overall process.

\begin{algorithm}
	%\textsl{}\setstretch{1.8}
	\renewcommand{\algorithmicrequire}{\textbf{Input:}}
	\renewcommand{\algorithmicensure}{\textbf{Output:}}
	\caption{Code-Aided Channel Estimation Algorithm}
	\label{alg1}
	\begin{algorithmic}[1]
		\REQUIRE the received vector $\boldsymbol{y}$, parity check matrix $H$, the LMMSE channel estimation result $\hat{\boldsymbol{H}}$
		\STATE $\tilde{\boldsymbol{H}} \leftarrow \hat{\boldsymbol{H}}$, $\tilde{\boldsymbol{H}}$ is the code-aided channel estimation
		\REPEAT
        \STATE  $\Delta\boldsymbol{H}\leftarrow\boldsymbol{0}_{N_r\times N_t}$
        \STATE obtain the LLRs of every bit given $\tilde{\boldsymbol{H}}$ via \eqref{ml}
		\FOR{$i=1$ to $N_r$}
		\FOR{$j=1$ to $N_t$}
		\STATE obtain the LLRs of every bit given $\Delta\boldsymbol{H},h$ via \eqref{LLRs}
		\STATE obtain $L_{\Pi_M}(h)$ via Algorithm~\ref{cal} 
		\STATE find $h_\text{max} = \underset{h}{\arg\max}L_{\Pi_M}(h)$
		\STATE $\Delta\boldsymbol{H} \leftarrow \Delta\boldsymbol{H} + h_\text{max} \boldsymbol{E}_{i,j}$ and update $\boldsymbol{p}_i$ and $\boldsymbol{q}_i$
		\ENDFOR
		\ENDFOR
		\STATE $\tilde{\boldsymbol{H}} \leftarrow \tilde{\boldsymbol{H}} + \Delta\boldsymbol{H}$ 
		\UNTIL{$\Vert\Delta\boldsymbol{H}\Vert_{F} \leq \varepsilon$}
		\ENSURE the code-aided channel estimation $\tilde{\boldsymbol{H}}$
	\end{algorithmic}  
\end{algorithm}

When optimizing the element at the $r$th row and $c$th column of $\tilde{\boldsymbol{H}}$, we first set
$$\Delta\boldsymbol{H}^{\prime}(h) = \Delta\boldsymbol{H} + h \boldsymbol{E}_{r,c}$$
and then get the \acp{LLR} of every bit by $\boldsymbol{y},\boldsymbol{H}^{\prime}$. The value of $h\in\mathbb{C}$ is enumerated to maximize $L_{\Pi_M}$. The step size is set to $b\sigma^2/n_p$ where $b$ is a parameter to balance the accuracy and computational complexity. The reason using this step size is that  $\Vert\hat{\boldsymbol{H}}-\boldsymbol{H}\Vert_{F}^2$ is approximately proportional to $\sigma^2/n_p$.

We assume that $\Vert\hat{\boldsymbol{H}}-\tilde{\boldsymbol{H}}\Vert_{F}$ is small. So the value of $\Vert\Delta\boldsymbol{H}\Vert_{F}$ is small in every iteration. Therefore, we assume that for $i=1,2,\ldots,N$, 
$$\boldsymbol{x}_i^1=\underset{\boldsymbol{x}\in Q_i^1}{\arg\min}\Vert\boldsymbol{y}- \tilde{\boldsymbol{H}}\boldsymbol{x}\Vert^2,\boldsymbol{x}_i^0=\underset{\boldsymbol{x}\in Q_i^0}{\arg\min}\Vert\boldsymbol{y}-\tilde{\boldsymbol{H}}\boldsymbol{x}\Vert^2$$
in \eqref{ml} are the same as the ones when $\tilde{\boldsymbol{H}}$ replaced by $\boldsymbol{H}^{\prime} = \tilde{\boldsymbol{H}} + \Delta\boldsymbol{H}^{\prime}$.

With given $h$ and $\{L_i\}_{i=1}^N$ under $\tilde{\boldsymbol{H}}+\Delta\boldsymbol{H}$, the \acp{LLR} under $\tilde{\boldsymbol{H}}+\Delta\boldsymbol{H}^{\prime}(h)$ is approximated as
\begin{equation}
	\begin{aligned}
		L_i^{\prime} \approx& L_i + \frac{1}{\sigma^2}\Re\left(\boldsymbol{p}_i^H h \boldsymbol{E}_{r,c}\boldsymbol{x}_i^0-\boldsymbol{q}_i^H h \boldsymbol{E}_{r,c}\boldsymbol{x}_i^1\right)\\
		&- \frac{1}{2\sigma^2}\left(\Vert h \boldsymbol{E}_{r,c}\boldsymbol{x}_i^0\Vert^2 - \Vert h \boldsymbol{E}_{r,c}\boldsymbol{x}_i^1\Vert^2\right)\\
		=& L_i + \frac{1}{\sigma^2}\Re\left(h\left(p_{i,r}^H x_{i,c}^0-q_{i,r}^H x_{i,c}^1\right)\right)\\
		&- \frac{\left| h\right|^2}{2\sigma^2}\left(\left| x_{i,c}^0\right|^2-\left|x_{i,c}^1\right|^2\right),
	\end{aligned}
	\label{LLRs}
\end{equation}
where $\boldsymbol{p}_i = \boldsymbol{y}_i-(\tilde{\boldsymbol{H}}+\Delta\boldsymbol{H})\boldsymbol{x}_i^0$ and $\boldsymbol{q}_i = \boldsymbol{y}_i-(\tilde{\boldsymbol{H}}+\Delta\boldsymbol{H})\boldsymbol{x}_i^1$. The values of $\boldsymbol{x}_i^0$ and $\boldsymbol{x}_i^1$ are stored when executing the ML detector to acquire LLRs under $\tilde{\boldsymbol{H}}$. The values of $\boldsymbol{p}_i$ and $\boldsymbol{q}_i$ are updated for any $h$ with complexity $\mathcal{O}(N)$ since $h \boldsymbol{E}_{r,c}$ has only one nonzero element. Without this approximation, we have to execute MIMO detection every time we calculate $L_{\Pi_M}$. Now we only need to execute once in every iteration.

Let $L_{\Pi_M,0}$ denote the value of $L_{\Pi_M}$ under $\tilde{\boldsymbol{H}}+\Delta\boldsymbol{H}$, $L_{\Pi_M,0}(h)$ denote the value of $L_{\Pi_M}$ under $\tilde{\boldsymbol{H}}+\Delta\boldsymbol{H}^{\prime}(h)$, $L_{\Pi_M,1}$ denote the value of $L_{\Pi_M}$ under $\tilde{\boldsymbol{H}}+\Delta\boldsymbol{H}(h_{max})$. Notice that $L_{\Pi_M,0}(0) = L_{\Pi_M,0}$, so $L_{\Pi_M,1} \geq L_{\Pi_M,0}$. Therefore, $L_{\Pi_M}$ is non-decreasing during iteration. In addition, let 
$$P_1=\underset{\hat{\boldsymbol{H}}}{\max }P_{Z_{1}|\bar{\boldsymbol{Y}}_d,\bar{\boldsymbol{H}}}(0|\boldsymbol{Y}_d,\hat{\boldsymbol{H}})$$
and we have $$P_{\Pi_M}(0)<P_{Z_{1}|\bar{\boldsymbol{Y}}_d,\bar{\boldsymbol{H}}}(0|\boldsymbol{Y}_d,\boldsymbol{H}^\star)\leq P_1$$
from \eqref{prod} and $P_1<1$ from its definition. So, $L_{\Pi_M}$ has an upper bound $\ln\left(P_1/(1-P_1)\right)$. Based on these two points, The iteration in Algorithm~\ref{alg1} converges. Indeed, the Algorithm~\ref{alg1} converges to a local maximum of \eqref{prod}.

The computational complexity of the coordinate ascent algorithm can be expressed as $\mathcal{O}\left( N_1\left( N2^{kN_t} +N N_t N_r+
		\notag \right.\right.
		\\
		\left.\left.N_2MW\right)\right),$
where $N_1$ is the average iteration count of Algorithm~\ref{alg1} and $N_2$ is the average number of executing Algorithm~\ref{cal} in line 8 of Algorithm~\ref{alg1}. The first term of the computational complexity pertains to the MIMO detection, the second term is for the updates of $\boldsymbol{p}_i$ and $\boldsymbol{q}_i$ and the third term is the number of times Algorithm 1 is executed. Through simulation, we find that satisfying performance is achieved with limited $N_1$ $(<10)$. Therefore, the complexity of Algorithm~\ref{alg1} is acceptable. Furthermore, this complexity can be reduce via \ac{MIMO} detection with lower complexity, e.g., sphere decoding~\cite{sphere}.

Once the code-aided channel estimation $\tilde{\boldsymbol{H}}$ is obtained, we execute the BP decoder to obtain the decoded word $\hat{\boldsymbol{u}}$.

\begin{figure}[!h]
	\centering
	\includegraphics[width=0.45\textwidth]{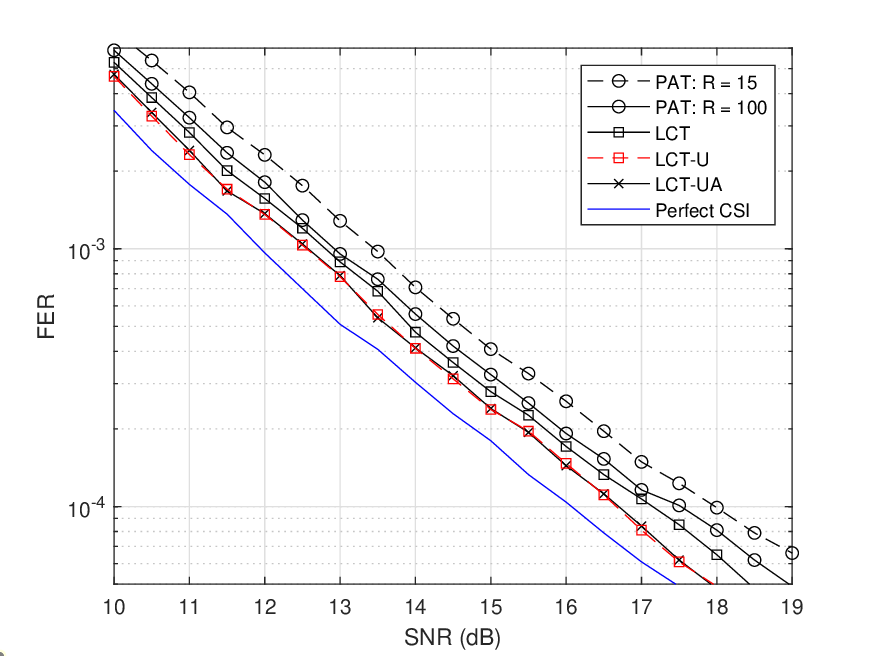}
	\caption{Performance of PAT and LCT with $\boldsymbol{H} \in \mathbb{C}^{2 \times 2}$. A $(192,96)$ LDPC code is adopted with QPSK such that the number $n_d$ of coded symbols is 48. The number $n_p$ of pilot symbols is set to $15$.}
	\label{eff}
	\vspace{-0.3cm}
\end{figure}

\begin{figure}[!t]
	\centering
	\includegraphics[width=0.45\textwidth]{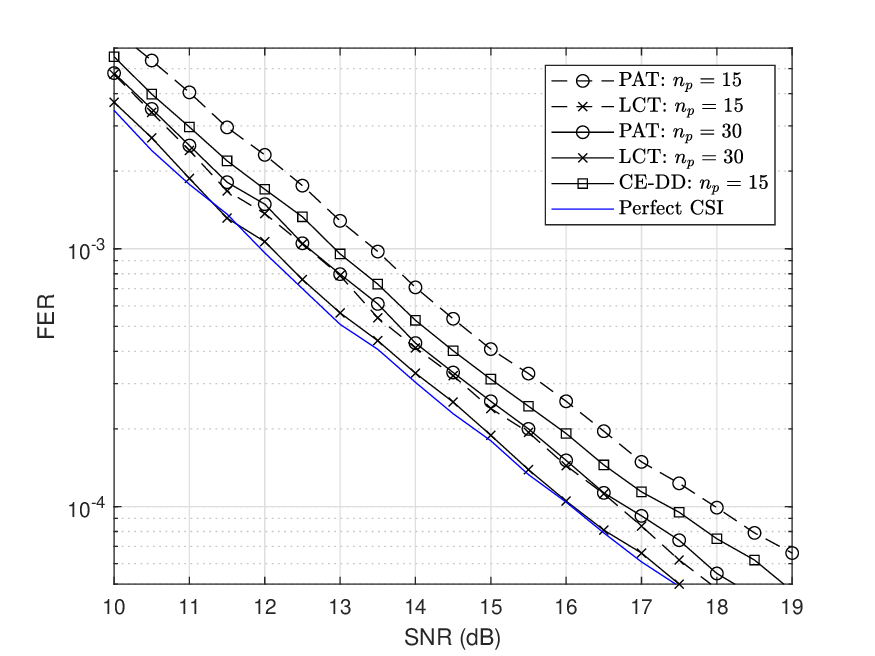}
	\caption{Performance of PAT and LCT with $\boldsymbol{H} \in \mathbb{C}^{2 \times 2}$. A $(192,96)$ LDPC code is adopted with QPSK such that the number $n_d$ of coded symbols is 48. The number $n_p$ of pilot symbols is set to $15$ and $30$.}
	\label{np}
	\vspace{-0.3cm}
\end{figure}

\section{Numerical Results}

In this section, we provide Monte Carlo simulation results and compare the performance of \ac{PAT} and \ac{LCT}. The SNR is expressed as ${E_sN_t}/(2\sigma^2 N_r)$, where $E_s$ is the energy per symbol. \Ac{FER} is chosen to be the performance criteria. We select the LDPC code in WiMAX with $N = 192$ and $K = 96$. Rayleigh block-fading channel with $2\times2$ \ac{MIMO} is considered in all the simulations, e.g., the entries of $\boldsymbol{H}$ are i.i.d. as $\mathcal{C N}\left(0,1\right)$. The maximum iteration count $R$ is set to 15 in all the simulations unless otherwise indicated. The performance is also compared to a coherent receiver which perfect \ac{CSI} is available for detection. The Monte Carlo simulation terminates when the number of trials reaches $10^6$ or the number of frame errors of coherent receiver with the perfect \ac{CSI} reaches 200.

\subsection{Performance}

This part shows the performance of the two methods. We adopt \ac{QPSK} and $n_p=15$ for all conditions. The \acp{FER} of different methods are shown in Fig.~\ref{eff}. The iteration count R is additionally set to $100$ in PAT to eliminate the influence of insufficient iteration count. \Ac{LCT} refers to Algorithm~\ref{alg1} using \eqref{ml} to obtain the LLR and Algorithm~\ref{cal}a, LCT-U refers to Algorithm~\ref{alg1} adopting \eqref{ml} and Algorithm~\ref{cal}, and LCT-UA stands for Algorithm~\ref{alg1} adopting Algorithm~\ref{cal} and the approximation of \ac{LLR} in \eqref{LLRs}.

As shown in Fig.~\ref{eff}, at an \ac{FER} $\approx 10^{-4}$, the LCT-UA and LCT-U outperform PAT with $R=100$ by about $0.8$~dB and \ac{PAT} with $R=15$ by about $1.3$~dB. They also perform within $0.6$~dB of the receiver with perfect \ac{CSI}, which proves the effectiveness of \ac{LCT}. Moreover, the benefit of updating $L_{C_{j}\left|\Pi_{i}\right.}$ via \eqref{update} is proven because LCT-U outperforms LCT by about $0.5$~dB. We can also find that LCT-UA and LCT-U exhibit identical, which means that the 
approximation in \eqref{LLRs} does not incur any performance penalty. Therefore, we adopt LCT-UA in the following simulations and denote it as LCT.

The gains may originate from a more precise value relation between \ac{LLR}s of every bit. ML detection with channel estimation $\tilde{\boldsymbol{H}}$ represents the constrains of \ac{LLR}s and Algorithm~\ref{alg1} provides the most probable \ac{LLR}s under these constrains.

\subsection{Impact of $n_p$}
Next, we consider the performance of \ac{LCT} with different values of $n_p$ to show that the performance gain persists with sufficient pilot symbols. We use \ac{QPSK} for all the simulations. The values of $n_p$ are $15$ and $30$. As shown in Fig.~\ref{np}, the performance of \ac{LCT} with $n_p=30$ still outperforms PAT with $n_p=30$ by about $0.7$~dB at an FER $\approx 10^{-4}$. At the same FER, \ac{LCT} performs within $0.1$~dB compared to the receiver with perfect \ac{CSI} when $n_p$ is $30$. The performance of LCT is also compared with CE-DD in \cite{4156391} for $n_p = 15$. The performance of \ac{LCT} outperforms CE-DD by about $0.5$~dB at an FER $\approx 10^{-4}$ for $n_p=15$.

\begin{figure}[!t]
	\centering
	\includegraphics[width=0.45\textwidth]{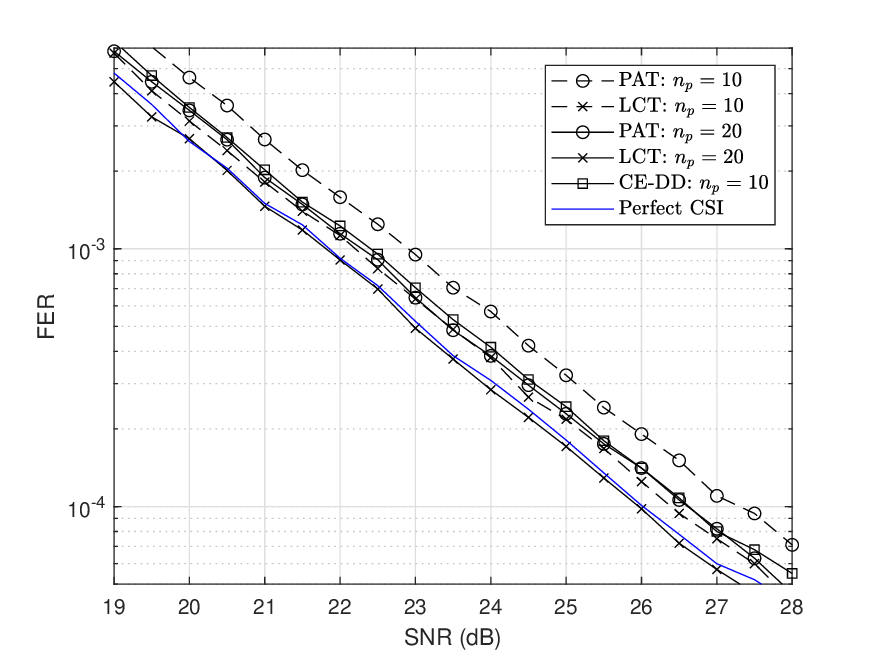}
	\caption{Performance of PAT and LCT with $\boldsymbol{H} \in\mathbb{C}^{2 \times 2}$. A $(192,96)$ LDPC code is used with $16$-QAM such that the number $n_d$ of coded symbols is 24. The number $n_p$ of pilot symbols is set to $10$ and $20$.}
	\label{fig_b2}
	\vspace{-0.3cm}
\end{figure}

\begin{table}[!t]
	\caption{The values of average iteration count of Algorithm~\ref{alg1}}
	\label{tab:table1}
	\centering
	\begin{tabular}{c c c|c c c}
		\hline
		\hline
		SNR (dB) & Modulation & $N_1$ & SNR (dB) & Modulation & $N_1$\\
		\hline
		10 & QPSK & 4.36 & 19 & 16-QAM & 3.15\\
		13 & QPSK & 5.84 & 22 & 16-QAM & 3.97\\
		16 & QPSK & 6.91 & 25 & 16-QAM & 4.64 \\
		19 & QPSK & 7.69 & 28 & 16-QAM & 5.08 \\	
		\hline
		\hline
	\end{tabular}
	\vspace{-0.3cm}
\end{table}

\subsection{Higher-order Modulation}
We next consider the performance of \ac{LCT} with $16$-QAM under different values of $n_p$. The values of $n_p$ are $10$ and $20$. As shown in Fig.~\ref{fig_b2}, the performance of \ac{LCT} outperforms \ac{PAT} by about $0.6$~dB when $n_p=10$ and about $0.9$~dB when $n_p=20$ at an FER $\approx 10^{-4}$. At the same FER, the performance of \ac{LCT} approaches closely that of the receiver with perfect CSI when $n_p$ is $20$. The performance of \ac{LCT} outperforms CE-DD by about $0.2$~dB at an FER $\approx 10^{-4}$ when $n_p=10$.

The values of $N_1$ under \ac{QPSK} and $16$-QAM are also given in Table \ref{tab:table1}. The values of $n_p$ for QPSK and $16$-QAM are $15$ and $10$, respectively. We can find that the values of $N_1$ are not large so the complexity of Algorithm~\ref{alg1} is acceptable.

\section{Conclusion}
Through introducing the parity-check constraints of an \ac{LDPC} code, an \ac{LCT} scheme was proposed to fine-tune the estimated \ac{CSI} in \ac{MIMO} system. The coordinate ascent algorithm was adopted to handle the exponential growth of solution space. Furthermore, an \ac{LLR} approximation method was proposed to reduce the times of \ac{MIMO} detection required by the algorithm. Simulation results showed that \ac{LCT} outperforms \ac{PAT} schemes with acceptable complexity.

%\section*{Acknowledgments}
%This should be a simple paragraph before the References to thank those individuals and institutions who have supported your work on this article.

%\vfill

\end{document}